\documentclass[10pt,conference]{IEEEtran}\IEEEoverridecommandlockouts
\usepackage{epsfig,graphicx,subfigure,psfrag,amsmath,cases}
\usepackage{latexsym,amssymb,amsmath,epsfig,subfigure,algorithm,mathtools}
\usepackage{algorithmic}
\usepackage{color}
\usepackage{url}
\usepackage{scrtime}
\usepackage{array}
\usepackage{bm}

\author{\authorblockN{Elena Boshkovska\authorrefmark{3},  Alexander Koelpin\authorrefmark{3}, Derrick Wing Kwan Ng\authorrefmark{1}, Nikola Zlatanov\authorrefmark{2}, and Robert Schober\authorrefmark{3}\thanks{Robert Schober is also with the University of British Columbia. This work was supported in part by the AvH Professorship Program of the Alexander von Humboldt Foundation.}}
Friedrich-Alexander-University Erlangen-N\"urnberg (FAU), Germany\authorrefmark{3}\\
The University of New South Wales, Australia\authorrefmark{1}\\
Monash University, Australia\authorrefmark{2} \vspace*{-3mm}
}

\title{\vspace*{-6mm}Robust Beamforming for
SWIPT Systems with Non-linear Energy Harvesting
Model\vspace*{-1mm}}

\date{\thistime,\,\today}

\newtheorem{Thm}{Theorem}
\newtheorem{Lem}{Lemma}

\newtheorem{T-Prob}{Transformed Problem}

\DeclareMathOperator{\Tr}{\mathrm{Tr}}
\DeclareMathOperator{\zero}{\mathbf{0}}
\DeclareMathOperator{\Rank}{\mathrm{Rank}}

\DeclareMathOperator{\vect}{\mathrm{vec}}
\DeclareMathOperator{\maxo}{\mathrm{maximize}}
\DeclareMathOperator{\mino}{\mathrm{minimize}}

\DeclareMathAlphabet\mathbfcal{OMS}{cmsy}{b}{n}

 \newcommand{\qed}{\hfill \ensuremath{\blacksquare}}

\newcommand{\abs}[1]{\lvert#1\rvert}

\newcommand{\norm}[1]{\lVert#1\rVert}
\newcolumntype{L}{>{\arraybackslash\raggedright}m{4cm}}
\begin{document}
\IEEEspecialpapernotice{(Invited Paper)}
\maketitle

\begin{abstract}
This paper investigates resource allocation for simultaneous wireless information and power transfer (SWIPT)
downlink  systems based on a non-linear energy harvesting model.  The resource allocation algorithm design is formulated as a non-convex optimization problem for the maximization of the total harvested power. The proposed problem formulation not only takes into account  imperfect channel state information (CSI) but also guarantees the quality-of-service (QoS) of information transfer.  A novel iterative algorithm is proposed to obtain the globally optimal solution of the considered non-convex optimization problem. In each iteration, a rank-constrained semidefinite program (SDP) is solved
optimally by SDP relaxation.  Simulation results  demonstrate the significant gains in harvested
power and the robustness against CSI imperfection  for the proposed optimal resource allocation, compared to a baseline scheme designed for perfect CSI and the conventional linear energy harvesting model.
\end{abstract}
\renewcommand{\baselinestretch}{0.90}

\section{Introduction}
\label{sect1}
The development of the Internet of Things (IoT)  has triggered an exponential  growth in the number of wireless communication devices  worldwide for applications such as  environmental monitoring, energy management, and safety management, etc. \cite{JR:IoT}. In particular, battery powered wireless sensor modules will be unobtrusively and invisibly integrated into clothing, walls, and vehicles, at locations which are inaccessible for wired recharging. The limited lifetime of wireless nodes creates a bottleneck for communication networks. As a result, wireless powered communication was proposed in the literature \cite{CN:Shannon_meets_tesla}\nocite{Krikidis2014,JR:QQ_WPC,JR:WIPT_fullpaper_OFDMA,JR:MIMO_WIPT,JR:SWIPT_imperfect_CSI,JR:Kwan_secure_imperfect,JR:non_linear_model,DBLP:journals/corr/Boshkovska16}--\cite{JR:non_linear_model}. Specifically,  wireless communication devices harvest energy from ambient propagating electromagnetic (EM) waves  in radio frequency (RF) for extending their lifetimes and supporting the energy consumption required for future information transmission. Besides,  wireless  channels are broadcast channels which facilitates
the possibility of simultaneous wireless information and power transfer
(SWIPT) leading to a new paradigm in wireless communication system design.


Recently,  the literature has focused on resource allocation algorithm designs that improve the efficiency of various  SWIPT systems \cite{JR:QQ_WPC}--\cite{JR:non_linear_model}. In \cite{JR:QQ_WPC} and \cite{JR:WIPT_fullpaper_OFDMA},  resource allocation algorithms were studied for the maximization of the achievable energy efficiency of single-carrier and multi-carrier SWIPT networks, respectively. In \cite{JR:MIMO_WIPT}, by exploiting the extra degrees of freedom offered by multiple antennas, beamforming was proposed to maximize the total transferred wireless power. However, the results in
\cite{JR:QQ_WPC}--\cite{JR:MIMO_WIPT} were based on the overly optimistic assumption of perfect channel state information (CSI). On the other hand,
 beamforming designs for secure SWIPT networks with the consideration of imperfect CSI were investigated in \cite{JR:SWIPT_imperfect_CSI} and \cite{JR:Kwan_secure_imperfect} for different system settings. However,  in most of the literature  \cite{CN:Shannon_meets_tesla}--\cite{JR:Kwan_secure_imperfect}, resource allocation algorithms  were designed based on the assumption of a  linear energy harvesting model which was recently shown to be inaccurate and not capable of capturing the non-linear behaviour of RF energy harvesting circuits \cite{JR:non_linear_model}. Unfortunately, resource allocation algorithms designed for the over simplified linear energy harvesting  model may lead to resource allocation mismatches resulting in severe performance degradation.
Motivated by the aforementioned prior works, this paper studies the optimal resource allocation algorithm design for SWIPT systems based on a non-linear energy harvesting model, which provides efficient SWIPT despite the imperfect CSI knowledge.

\textbf{Notation:}
 In this paper, we adopt the following notations. $\mathbf{A}^H$, $\Tr(\mathbf{A})$, and $\Rank(\mathbf{A})$ represent the  Hermitian transpose, trace, and rank of  matrix $\mathbf{A}$; $\mathbf{A}\succeq \mathbf{0}$ indicates that $\mathbf{A}$ is a  positive semidefinite matrix; matrix $\mathbf{I}_{N}$
denotes an $N\times N$ identity matrix.  $\vect(\mathbf{A})$ denotes the vectorization of matrix $\mathbf{A}$.
$\mathbf{A}\otimes \mathbf{B}$ denotes the Kronecker product of matrices $\mathbf{A}$ and $ \mathbf{B}$. $[\mathbf{B}]_{a:b,c:d}$ returns a submatrix of $\mathbf{B}$ including the $a$-th to the $b$-th rows and the $c$-th to the $d$-th columns of $\mathbf{B}$. $[\mathbf{q}]_{m:n}$  returns a vector with the $m$-th to the $n$-th elements of vector $\mathbf{q}$. A complex Gaussian random  vector with mean vector $\bm\mu$ and covariance matrix
$\bm\Sigma$ is denoted by ${\cal CN}(\bm\mu,\bm\Sigma)$, and $\sim$ means
``distributed as".
$\mathbb{C}^{N\times M}$ denotes the space of all $N\times M$ matrices with complex entries.
$\mathbb{H}^N$ represents the set of all $N$-by-$N$ complex Hermitian matrices. $\cal E\{\cdot\}$ denotes statistical expectation.  $\abs{\cdot}$, $\norm{\cdot}$, and $\norm{\cdot}_F$ denote the absolute value of a complex scalar, the Euclidean norm, and the Frobenius norm of a vector/matrix, respectively; $\mathrm{Re}\{\cdot\}$ denotes the real part of an input complex number.
\section{System Model}
In this section, we define the  channel and energy harvesting models adopted for resource allocation algorithm design.

\begin{figure}[t]
\centering
\includegraphics[width=3.5 in]{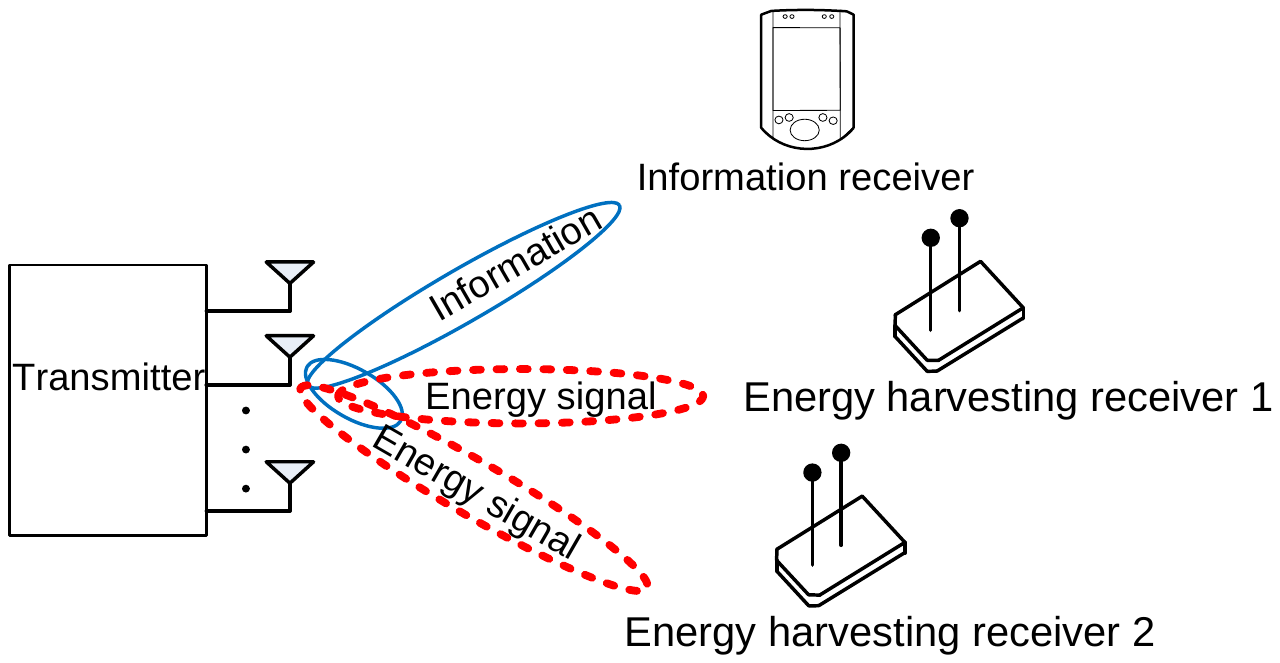}
\caption{A downlink SWIPT communication system with an information receiver  and $J=2$ energy harvesting receivers (ERs).}
\label{fig:system_model}
\end{figure}

\subsection{Channel Model}
We consider a flat fading channel for downlink SWIPT systems.
The system consists of a transmitter, an  information receiver,
 and $J$ energy harvesting receivers, cf. Figure \ref{fig:system_model}. The transmitter  is equipped with $N_{\mathrm{T}}\geq 1$ antennas.  The information receiver is a single-antenna device  and each  energy harvesting receiver is equipped with $N_{\mathrm{R}}\geq 1$ receive antennas to facilitate energy harvesting. In each time slot, the received signals at the information receiver and  energy harvesting receiver $j\in\{1,\ldots, J\}$ are given by
\begin{eqnarray}
y&=&\mathbf{h}^H(\mathbf{w}s+\mathbf{v}) +n,\,\, \mbox{and}\\
\mathbf{y}_{\mathrm{ER}_j}&=&\mathbf{G}_j^H(\mathbf{w}s+\mathbf{v})+\mathbf{n}_{\mathrm{ER}_j},\,\,  \forall j\in\{1,\dots,J\},
\end{eqnarray}respectively, where $s\in\mathbb{C}$ and $\mathbf{w}\in\mathbb{C}^{N_{\mathrm{T}}\times1}$  are the data symbol  and  the information beamforming vector, respectively.  Without loss of generality, we assume that ${\cal E}\{\abs{s}^2\}=1$.  $\mathbf{v}\in\mathbb{C}^{N_{\mathrm{T}}\times 1}$ is an energy signal vector generated by the transmitter to facilitate efficient wireless power transfer.  $\mathbf{v}$ is modeled as a complex Gaussian random vector with $\mathbf{v}\sim {\cal CN}(\mathbf{0}, \mathbf{V})$.
 The channel vector between the transmitter and the information receiver is denoted by $\mathbf{h}\in\mathbb{C}^{N_{\mathrm{T}}\times1}$ and the channel matrix between the transmitter and  energy harvesting receiver $j$  is denoted by $\mathbf{G}_j\in\mathbb{C}^{N_{\mathrm{T}}\times N_{\mathrm{R}}}$. $n\sim{\cal CN}(0,\sigma_{\mathrm{s}}^2)$ and $\mathbf{n}_{\mathrm{ER}_j}\sim{\cal CN}(\zero,\sigma_{\mathrm{s}}^2\mathbf{I}_{N_{\mathrm{R}}})$ are the additive white Gaussian noises (AWGN) at  the  information receiver and energy harvesting receiver $j$, respectively, where $\sigma_{\mathrm{s}}^2$ denotes the noise power at the receiver.
\subsection{Non-linear Energy Harvesting Model}
 The total received RF power at energy harvesting receiver $j$  is given by \begin{eqnarray}\label{eqn:linear_model}
P_{\mathrm{ER}_j}=\Tr\Big((\mathbf{w}\mathbf{w}^H +\mathbf{V})\mathbf{G}_j\mathbf{G}_j^H\Big).
\end{eqnarray}
In practice, an energy harvesting circuit \cite{JR:Energy_harvesting_circuit}\nocite{JR:EH_measurement_1}--\cite{CN:EH_measurement_2}  is equipped at the energy harvesting receiver which is used to convert the received RF power into direct current (DC) power for future use. Yet,  practical energy harvesting circuits introduce various non-linearities into the end-to-end wireless power transfer.  In this paper, we adopt a newly proposed non-linear parametric energy harvesting model from \cite{JR:non_linear_model} for resource allocation algorithm design.  In particular, based on experimental results, it has been verified that the parametric non-linear model proposed in \cite{JR:non_linear_model}  is able to accurately capture the dynamics of the RF energy conversion efficiency for different input power levels and the joint effects of the non-linear phenomena caused by hardware imperfections. The total harvested power at energy harvesting receiver $j$, $\Phi_{\mathrm{ER}_j}$, is modelled as:
 \begin{eqnarray}\label{eqn:EH_non_linear}
 \hspace*{-5mm}\Phi_{\mathrm{ER}_j}\hspace*{-2mm}&=&\hspace*{-2mm}
 \frac{[\Psi_{\mathrm{ER}_j}
 \hspace*{-0.5mm}- \hspace*{-0.5mm}M_j\Omega_j]}{1-\Omega_j},\, \Omega_j=\frac{1}{1+\exp(a_jb_j)},\\
 \hspace*{-5mm}\mbox{where}\,\,\Psi_{\mathrm{ER}_j}\hspace*{-2mm}&=&\hspace*{-2mm} \frac{M_j}{1+\exp\Big(\hspace*{-0.5mm}-a_j(P_{\mathrm{ER}_j}-\hspace*{-0.5mm}b_j)\Big)}
  \end{eqnarray}
 is a logistic function which has the received RF power, $P_{\mathrm{ER}_j}$,  as the input. $M_j$ is a constant denoting the maximum harvested power at energy harvesting receiver $j$  when the energy harvesting circuit is saturated because of exceedingly large input power. Parameters $a_j$ and $b_j$ are constants which capture the joint effects of  resistance, capacitance, and circuit sensitivity. Specifically, $a_j$ reflects the non-linear charging rate with respect to the input power and $b_j$ is related to the minimum turn-on voltage of an energy harvesting circuit.   In practice,  parameters $a_j$, $b_j$, and $M_j$ of the proposed model in \eqref{eqn:EH_non_linear}
can be easily found using a standard curve fitting algorithm for a given energy harvesting hardware circuit.

\subsection{Channel State Information}
 In this paper, we assume that the transmitter has imperfect CSI. To capture the impact of the CSI imperfection on  resource allocation design, a deterministic model \cite{JR:Robust_error_models1,JR:SWIPT_DAS} is adopted. The  CSI of the links between the transmitter and the information receiver as well as energy harvesting receiver $j$  can be modelled as:
\begin{eqnarray}\label{eqn:outdated_CSI}
\mathbf{h}&=&\mathbf{\widehat h} + \Delta\mathbf{h},\,    \mbox{   and}\\
{\bm\Lambda }&\triangleq& \Big\{\Delta\mathbf{h}\in \mathbb{C}^{N_{\mathrm{T}}\times 1}  :\norm{\Delta\mathbf{h}}_2^2 \le \rho^2\Big\}, \label{eqn:outdated_CSI-set2}\\
\mathbf{G}_j&=&\mathbf{\widehat G}_j + \Delta\mathbf{G}_j,\,   \forall j\in\{1,\ldots,J\}, \\
{\bm\Xi }_j&\triangleq& \Big\{\Delta\mathbf{G}_j\in \mathbb{C}^{N_{\mathrm{T}}\times {N_{\mathrm{R}}}}  :\norm{\Delta\mathbf{G}_j}_F^2 \le \upsilon_j^2\Big\},\forall j,\label{eqn:outdated_CSI-set}
\end{eqnarray}
respectively, where $\mathbf{\widehat h}$ and $\mathbf{\widehat G}_j$ are the estimates of channel vector $\mathbf{ h}$ and matrix $\mathbf{G}_j$, respectively. $\Delta\mathbf{h}$ and $\Delta\mathbf{G}_j$  represent the channel uncertainty due to channel estimation errors. In  \eqref{eqn:outdated_CSI-set2} and (\ref{eqn:outdated_CSI-set}), sets  ${\bm\Lambda }$ and ${\bm\Xi }_j$  define continuous spaces spanned by all possible channel uncertainties,  respectively.  Constants $\rho$ and  $\upsilon_j$  denote the maximum value of the norm of the CSI estimation error vector   $ \Delta\mathbf{h}$ and matrix $ \Delta\mathbf{G}_j$, respectively.

\section{Problem Formulation and Solution}
The  considered system design objective is to maximize the total harvested power  while providing QoS for reliable communication with the consideration of imperfect CSI. The resource allocation
algorithm design is formulated as the following optimization
problem\footnote{In the sequel,  since $\Omega_j$ does not affect the design of the optimal resource allocation policy, with a slight abuse of notation,  we will directly use $\Psi_{\mathrm{ER}_j}$ to represent the harvested power at ER $j$ for simplicity  of presentation.}:
\begin{eqnarray}\label{eqn:TP_maximization}
\underset{\mathbf{V}\in \mathbb{H}^{N_{\mathrm{T}}},\mathbf{w}}{\maxo}\,\, \hspace*{-2mm}&& \sum_{j=1}^J \min_{\Delta\mathbf{G}_j\in{\bm\Xi }_j }\Psi_{\mathrm{ER}_j}\\
\hspace*{-2mm}\mathrm{subject\,\,to}\,\, &&\mathrm{C1}:\,\,\norm{\mathbf{w}}_2^2 + \Tr(\mathbf{V})\leq P_{\mathrm{max}},\notag\\
\hspace*{-2mm}&&\hspace*{-10mm}\mathrm{C2}:\,\, \min_{\Delta\mathbf{h}\in{\bm\Lambda } }\frac{\mathbf{w}^H\mathbf{H}\mathbf{w}}
{ \Tr(\mathbf{V}\mathbf{H})+\sigma_{\mathrm{s}}^2} \geq \Gamma_{\mathrm{req}},\quad \mathrm{C3}:\,\, \mathbf{V}\succeq \zero,\notag
\end{eqnarray}
where $\mathbf{H}=\mathbf{h}\mathbf{h}^H$. Constants $P_{\max}$ and $\Gamma_{\mathrm{req}}$  in constraints C1 and C2  are the maximum transmit power and the minimum required signal-to-interference-plus-noise ratio (SINR) at the information receiver, respectively.  C3 and $\mathbf{V}\in \mathbb{H}^{N_{\mathrm{T}}}$ constrain matrix $\mathbf{V}$  to be a positive semidefinite Hermitian matrix. It can be observed that the objective function in \eqref{eqn:TP_maximization} is a non-convex function and there are infinitely many inequality constraints in C2. In order to obtain a tractable solution, we first transform the non-convex objective function into an equivalent objective function in subtractive form via the following theorem.


\begin{Thm}\label{Thm:1}
Suppose $\{\mathbf{w}^*,\mathbf{V}^*\}$ is the optimal solution to   \eqref{eqn:TP_maximization}, then there exist two vectors ${\bm \mu}^*=[\mu_1^*,\ldots,\mu_J^*]$ and ${\bm \beta}^*=[\beta_1^*,\ldots,\beta_J^*]$ such that $\{\mathbf{w}^*,\mathbf{V}^*\}$ is an optimal solution to the following optimization problem
\begin{equation} \label{eqn:transformed}
\underset{\mathbf{V}^*\in \mathbb{H}^{N_{\mathrm{T}}},\mathbf{w}^*\in {\cal F}}\maxo\,\sum_{j=1}^J \mu_j^*\Big[\hspace*{-0.5mm}M_j\hspace*{-0.5mm}- \hspace*{-0.5mm} \beta_j^*\Big(1+\exp\big(\hspace*{-0.5mm}-\hspace*{-0.5mm}a_j(P_{\mathrm{ER}_j}\hspace*{-0.5mm}-\hspace*{-0.5mm}b_j)\big)\Big)\hspace*{-0.5mm}\Big],
\end{equation}
where $\cal F$ is the feasible solution set of \eqref{eqn:TP_maximization}. Besides, $\{\mathbf{w}^*,\mathbf{V}^*\}$  also satisfies the following system of equations:
\begin{eqnarray} \label{eqn:conditions1}
\beta_j^*\Big(1+\exp\big(\hspace*{-0.5mm}-\hspace*{-0.5mm}a_j(P_{\mathrm{ER}_j}^*\hspace*{-0.5mm}-\hspace*{-0.5mm}b_j)\big)\Big)-M_j&=&0,\\
\mu_j^*\Big(1+\exp\big(\hspace*{-0.5mm}-\hspace*{-0.5mm}a_j(P_{\mathrm{ER}_j}^*\hspace*{-0.5mm}-\hspace*{-0.5mm}b_j)\big)\Big)-1&=&0, \label{eqn:conditions2}
\end{eqnarray}
and $P_{\mathrm{ER}_j}^*=\Tr\Big((\mathbf{w}^*(\mathbf{w^*})^H +\mathbf{V}^*)\mathbf{G}_j\mathbf{G}_j^H\Big)$.
\end{Thm}

\,\,\emph{Proof:} Please refer to \cite{JR:sum_of_ratios} for a proof of  Theorem 1.

As a result,  for the maximization problem in \eqref{eqn:TP_maximization},
there exists an equivalent parametric optimization problem with an objective function in subtractive form and
 both problems have the same optimal solution $\{\mathbf{w}^*,\mathbf{V}^*\}$.
More importantly,  the optimization problem with an objective function in subtractive form can be solved by an iterative algorithm  consisting of two nested loops \cite{JR:sum_of_ratios}.
  In the inner loop, we solve the optimization in \eqref{eqn:transformed} for given $(\bm{\mu},\bm{\beta})$.  Then, in the outer loop,  we find the optimal $(\bm{\mu}^*,\bm{\beta}^*)$ satisfying the system of equations in \eqref{eqn:conditions1} and \eqref{eqn:conditions2}, cf. algorithm in Table \ref{table:algorithm}.
 \subsection{Solution of the Inner Loop Problem}
In each iteration, i.e., line 3 of the algorithm in Table I,  we solve the following inner loop non-convex optimization problem:
\begin{eqnarray}\label{eqn:TP_maximization-transfomred}\notag
\underset{\mathbf{W},\mathbf{V}\in \mathbb{H}^{N_{\mathrm{T}}},\bm\tau}{\maxo}\,\, \hspace*{-1mm}&&\hspace*{-2mm}\sum_{j=1}^J \mu_j^*\Big[\hspace*{-0.5mm}M_j\hspace*{-0.5mm}- \hspace*{-0.5mm} \beta_j^*\Big(1+\exp\big(\hspace*{-0.5mm}-\hspace*{-0.5mm}a_j(\tau_j\hspace*{-0.5mm}-\hspace*{-0.5mm}b_j)\big)\Big)\hspace*{-0.5mm}\Big]\\
\hspace*{-1mm}\mathrm{subject\,\,to}\,\, &&\hspace*{-2mm}\mathrm{C1}:\,\,\Tr(\mathbf{W+V})\leq P_{\mathrm{max}},\notag\\
\hspace*{-1mm}&&\hspace*{-2mm}\mathrm{C2}:\,\, \min_{\Delta\mathbf{h}\in{\bm\Lambda } }\frac{\Tr(\mathbf{W}\mathbf{H})}
{ \Tr(\mathbf{V}\mathbf{H})+\sigma_{\mathrm{s}}^2} \geq \Gamma_{\mathrm{req}},\notag\\
\hspace*{-1mm}&&\hspace*{-2mm}\mathrm{C4}:\,\, \min_{\Delta\mathbf{G}_j\in{\bm\Xi }_j }\Tr\Big((\mathbf{W}+\mathbf{V})\mathbf{G}_j\mathbf{G}_j^H\Big) \geq \tau_j,\notag \forall j,\\
\hspace*{-5mm}&&\hspace*{-23mm}\mathrm{C3}:\,\, \mathbf{V}\succeq \zero,\,\,\mathrm{C5}:\,\, \Rank(\mathbf{W})=1,\,\, \mathrm{C6}:\,\, \mathbf{W}\succeq\zero,
\end{eqnarray}
where $\mathbf{W}=\mathbf{w}\mathbf{w}^H$ is a new optimization variables matrix  and $\bm\tau=[\tau_1,\tau_2,\ldots,\tau_J]$ is a vector of auxiliary optimization variables.
\begin{table}[t]\caption{}\label{table:algorithm}
\vspace*{-6mm}
\begin{algorithm} [H]                    
\renewcommand\thealgorithm{}
\caption{Iterative Resource Allocation Algorithm }          

\label{alg1}                           
\begin{algorithmic} [1]
\STATE Initialize the maximum number of iterations $L_{\max}$, iteration index $n=0$, $\bm\mu$, and $\bm \beta$

\REPEAT [Outer Loop]
\STATE Solve the inner loop problem in \eqref{eqn:TP_maximization-transfomred} via SDP relaxation for
 given $(\bm\mu^n,\bm\beta^n)$ and obtain the intermediate beamformer $\mathbf{w}'$ and energy signal covariance matrix $\mathbf{V}'$
\IF { \eqref{eqn:convergence_condition} is satisfied} \RETURN
Optimal beamformer $\mathbf{w}^*=\mathbf{w}'$ and energy signal covariance matrix $\mathbf{V}^*=\mathbf{V}'$
 \ELSE \STATE
Update $\bm \mu$ and $\bm\beta$ according to \eqref{eqn:update_beta} and $n=n+1$
 \ENDIF
 \UNTIL{\eqref{eqn:convergence_condition} is satisfied $\,$or $n=L_{\max}$}

\end{algorithmic}
\end{algorithm}\vspace*{-10mm}
\end{table}To further facilitate the solution, we transform constraints C2 and C4 into linear matrix inequalities (LMIs) using the following lemma:
 \begin{Lem}[S-Procedure \cite{book:convex}] Let a function $f_m(\mathbf{x}),m\in\{1,2\},\mathbf{x}\in \mathbb{C}^{N\times 1},$ be defined as
\begin{eqnarray}\label{eqn:S-procedure}
f_m(\mathbf{x})=\mathbf{x}^H\mathbf{A}_m\mathbf{x}+2 \mathrm{Re} \{\mathbf{b}_m^H\mathbf{x}\}+c_m,
\end{eqnarray}
where $\mathbf{A}_m\in\mathbb{H}^N$, $\mathbf{b}_m\in\mathbb{C}^{N\times 1}$, and $c_m\in\mathbb{R}$. Then, the implication $f_1(\mathbf{x})\le 0\Rightarrow f_2(\mathbf{x})\le 0$  holds if and only if there exists a $\delta\ge 0$ such that
\begin{eqnarray}\delta
\begin{bmatrix}
       \mathbf{A}_1 & \mathbf{b}_1          \\
       \mathbf{b}_1^H & c_1           \\
           \end{bmatrix} -\begin{bmatrix}
       \mathbf{A}_2 & \mathbf{b}_2          \\
       \mathbf{b}_2^H & c_2           \\
           \end{bmatrix}          \succeq \zero,
\end{eqnarray}
provided that there exists a point $\mathbf{\hat{x}}$ such that $f_m(\mathbf{\hat{x}})<0$.
\end{Lem}

Exploiting Lemma 1,  the original constraint C2 holds if and only if there exists a $\delta\ge 0$,  such that the following  LMI constraint holds:
\begin{eqnarray}\label{eqn:LMI_C2}
&&\mbox{C2: }\mathbf{S}_{\mathrm{C}_{2}}\Big(\mathbf{W},\mathbf{V},\delta\Big)\\
&&\hspace*{-6.5mm}=
          \begin{bmatrix}
       \delta\mathbf{I}_{N_\mathrm{T}}  & \hspace*{-1mm}\zero        \\
       \zero     & \hspace*{-1mm}-\delta\rho^2 -\Gamma_{\mathrm{req}} \sigma_{\mathrm{s}}^2      \\
           \end{bmatrix} \hspace*{-0.5mm}+\hspace*{-0.5mm}\mathbf{U}_{\mathbf{\hat h}}^H(\mathbf{W}-\Gamma_{\mathrm{req}}\mathbf{V})\mathbf{U}_{\mathbf{\hat h}} \succeq \mathbf{0}, \notag
\end{eqnarray}
where $\mathbf{U}_{\mathbf{\hat h}}=\Big[\mathbf{I}_{N_{\mathrm{T}}}\quad\mathbf{\hat h}\Big]$. Similarly,  constraint C4 can be equivalently written as
\begin{eqnarray}\label{eqn:LMI_C4}&&
\mbox{C4: }\mathbf{S}_{\mathrm{C}_{4_j}}\Big(\mathbf{W},\mathbf{V}, \bm{\nu},\bm{\tau}\Big)\\
&=&\notag
           \begin{bmatrix}
       \nu_j\mathbf{I}_{N_{\mathrm{T}}N_{\mathrm{R}}}& \zero          \\
        \zero
        & -\tau_j-\nu_j\upsilon_j^2       \\
           \end{bmatrix}+ \mathbf{U}_{\widetilde{\mathbf{g}}_j}^H(\mathbfcal{ W}+\mathbfcal{V})\mathbf{U}_{\widetilde{\mathbf{g}}_j}\succeq \mathbf{0}, \forall j,\notag
\end{eqnarray}
for $\nu_j\ge 0, j\in\{1,\ldots,M\}$, $\mathbfcal{ W}=\mathbf{I}_{N_\mathrm{R}} \otimes \mathbf{W} $, $\mathbfcal{ V}=\mathbf{I}_{N_\mathrm{R}} \otimes \mathbf{V} $, $\mathbf{U}_{\widetilde{\mathbf{g}}_j}=[\mathbf{I}_{N_{\mathrm{T}}N_{\mathrm{R}}}\quad \widetilde{\mathbf{g}}_j]$, and $\widetilde{\mathbf{g}}_j=\vect({\mathbf{\hat G}}_j)$ . Then, the optimization problem can be equivalently written as
\begin{eqnarray}\label{eqn:TP_minimization}\notag
\underset{\underset{\bm \tau, \bm{\nu}}{\mathbf{W},\mathbf{V}\in \mathbb{H}^{N_{\mathrm{T}}}, \delta,}}{\maxo}\,\, \hspace*{-1mm}&&\hspace*{-2mm} \sum_{j=1}^J \mu_j^*\Big[\hspace*{-0.5mm}M_j\hspace*{-0.5mm}- \hspace*{-0.5mm} \beta_j^*\Big(1+\exp\big(\hspace*{-0.5mm}-\hspace*{-0.5mm}a_j(\tau_j\hspace*{-0.5mm}-\hspace*{-0.5mm}b_j)\big)\Big)\hspace*{-0.5mm}\Big]\\
\hspace*{0mm}\mathrm{subject\,\,to}\,\, &&\hspace*{-2mm}\mathrm{C1, C3,  C6},\notag\\
\hspace*{-0mm}&&\hspace*{-25mm}\mathrm{C2}:\, \mathbf{S}_{\mathrm{C}_{2}}\Big(\mathbf{W},\mathbf{V},\delta\Big) \succeq \zero,\,\, \mathrm{C4}:\,\mathbf{S}_{\mathrm{C}_{4_j}}\Big(\mathbf{W},\mathbf{V}, \bm{\nu},\bm{\tau}\Big) \succeq \zero,\notag \forall j,\\
\hspace*{-5mm}&&\hspace*{-25mm}\mathrm{C5}:\, \Rank(\mathbf{W})=1,\hspace*{8.5mm}\mathrm{C7}:\, \nu_j,\delta\geq 0,
\end{eqnarray}
where  $\delta$ and $\bm\nu=\{ \nu_1,\ldots,\nu_j,\ldots,\nu_J\}$  are the non-negative auxiliary optimization variables introduced by Lemma 1 for handling constraints C2 and C4, respectively.  We note that constraints C2 and C4 involve only a finite number of constraints which facilitates the resource allocation algorithm design.
The remaining obstacle in solving the considered optimization problem is the combinatorial rank constraint C5. We adopt the semidefinite programming (SDP) relaxation by removing constraint C5 from the problem formulation. As a result, the rank constraint relaxed problem becomes a standard convex optimization problem and can be solved efficiently by numerical solvers such as CVX \cite{website:CVX}. Yet, the constraint relaxation may not be tight if $\Rank(\mathbf{W}) > 1$ occurs. Therefore,   we reveal the tightness of the adopted SDP relaxation in \eqref{eqn:TP_maximization} in the following theorem.
\begin{Thm}\label{thm:rankone}
Assuming the considered problem is feasible for $\Gamma_{\mathrm{req}}>0$, a rank-one solution of \eqref{eqn:TP_maximization} can always be constructed.
\end{Thm}

\,\,\emph{Proof:} Please refer to the Appendix.

In other words, \eqref{eqn:TP_maximization-transfomred} can be solved optimally. Hence, information beamforming is optimal for the maximization of the total
harvested power, despite the existence of imperfect CSI.

\subsection{Solution of the Outer Loop Problem}
Now, we present an iterative algorithm to  update $(\bm{\mu},\bm{\beta})$  for the outer loop problem via the damped iterative Newton method.
For notational simplicity, we define functions $\varphi_j(\beta_j)=\beta_j\Big(1+\exp\big(\hspace*{-0.5mm}-\hspace*{-0.5mm}a_j(P_{\mathrm{ER}_j}\hspace*{-0.5mm}-\hspace*{-0.5mm}b_j)\big)\Big)-M_j$
and $\varphi_{J+i}(\mu_i)=\mu_i\Big(1+\exp\big(\hspace*{-0.5mm}-\hspace*{-0.5mm}a_i(P_{\mathrm{ER}_i}\hspace*{-0.5mm}-\hspace*{-0.5mm}b_i)\big)\Big)-1$, $i\in\{1,\ldots,J\}$. It is shown in \cite{JR:sum_of_ratios} that the unique optimal solution  $(\bm{\mu}^*,\bm{\beta}^*)$ is obtained if and only if  $\bm\varphi(\bm \mu,  \bm\beta)=[\varphi_1,\varphi_2,\ldots,\varphi_{2J}]=\zero$.  Therefore, in the $n$-th iteration of the iterative algorithm, ${\bm \mu}^{n+1}$ and ${\bm \beta}^{n+1}$ can be updated as, respectively,
\begin{eqnarray}\label{eqn:update_beta}
\hspace*{-3.5mm}{\bm \mu}^{n+1}\hspace*{-2.5mm}&=&\hspace*{-2.5mm}{\bm \mu}^{n}+\zeta^n\mathbf{q}^n_{J+1:2J}\,\mbox{and}\,  {\bm \beta}^{n+1}={\bm \beta}^{n}+\zeta^n\mathbf{q}^n_{1:J},\\
\hspace*{-2.5mm}\mbox{where }\,\,\mathbf{q}^n\hspace*{-3.5mm}&=&\hspace*{-2.5mm}[\bm\varphi'(\bm{\mu},\bm{\beta})]^{-1}\bm\varphi(\bm{\mu},\bm{\beta})
\end{eqnarray}
 and $\bm\varphi'(\bm{\mu},\bm{\beta})$ is the Jacobian matrix of $\bm\varphi(\bm{\mu},\bm{\beta})$. $\zeta^n$ is the largest  $\varepsilon^l$ satisfying
\begin{eqnarray}\label{eqn:convergence_condition}
\norm{\bm\varphi\big({\bm \mu}^{n}\hspace*{-0.5mm}+\hspace*{-0.5mm}\varepsilon^l\mathbf{q}^n_{J\hspace*{-0.5mm}+\hspace*{-0.5mm}1:2J},{\bm \beta}^{n}\hspace*{-0.5mm}+\hspace*{-1.5mm}\varepsilon^l\mathbf{q}^n_{1:J}\big)}\leq (1-\eta\varepsilon^l)\norm{\bm\varphi(\bm{\mu},\bm{\beta})},
\end{eqnarray}
where $l\in\{1,2,\ldots\}$, $\varepsilon^l\in(0,1)$, and $\eta\in(0,1)$. The damped Newton method converges to the unique solution $(\bm{\mu}^*,\bm{\beta}^*)$  satisfying the system of equations \eqref{eqn:conditions1} and \eqref{eqn:conditions2}, cf. \cite{JR:sum_of_ratios}.

\section{Results}
In this section, we evaluate the system performance of the proposed optimal resource allocation via simulations. The important simulation parameters are listed in Table \ref{table:parameters}. We assume that the information receiver and the $J=10$ energy harvesting receivers  are located at $50$ meters and $10$ meters from the transmitter, respectively. The information receiver requires a minimum SINR of $10$ dB.  In the sequel, we define the normalized maximum  channel estimation errors of energy harvesting receiver $j$  and the information receiver  as  $\sigma_{\mathrm{est}_{G}}^2=\frac{\upsilon^2_j}{\norm{\mathbf{G}_j}^2_F},\forall j,$  and $\sigma_{\mathrm{est}_h}^2=\frac{\rho^2}{\norm{\mathbf{h}}^2_2}=5\%$. For the non-linear EH circuits, we set $M_j=20$ mW which corresponds to the maximum harvested power per energy harvesting receiver. Besides, we adopt $a_j=6400$ and $b_j=0.003$.  We solve the
optimization problem in \eqref{eqn:TP_maximization} and obtain the average system
performance by averaging over different channel realizations.

\begin{table}[t]\vspace*{-4mm}
\caption{Simulation Parameters} \label{table:parameters}
\centering
\begin{tabular}{ | L | l | } \hline
      Carrier center frequency                           & 915 MHz\\ \hline
      Bandwidth                                          & $200$ kHz \\ \hline 
      Transceiver  antenna gain                                     & 12 dBi \\ \hline
      Noise power                                        & $\sigma^2= -95$ dBm \\ \hline
      Transmitter-to-energy harvesting receiver fading distribution                                      & Rician with Rician factor $3$ dB \\
        \hline
\end{tabular}\vspace*{-4mm}
\end{table}

\begin{figure}[t]
 \centering\vspace*{-8mm}
\includegraphics[width=3.2in]{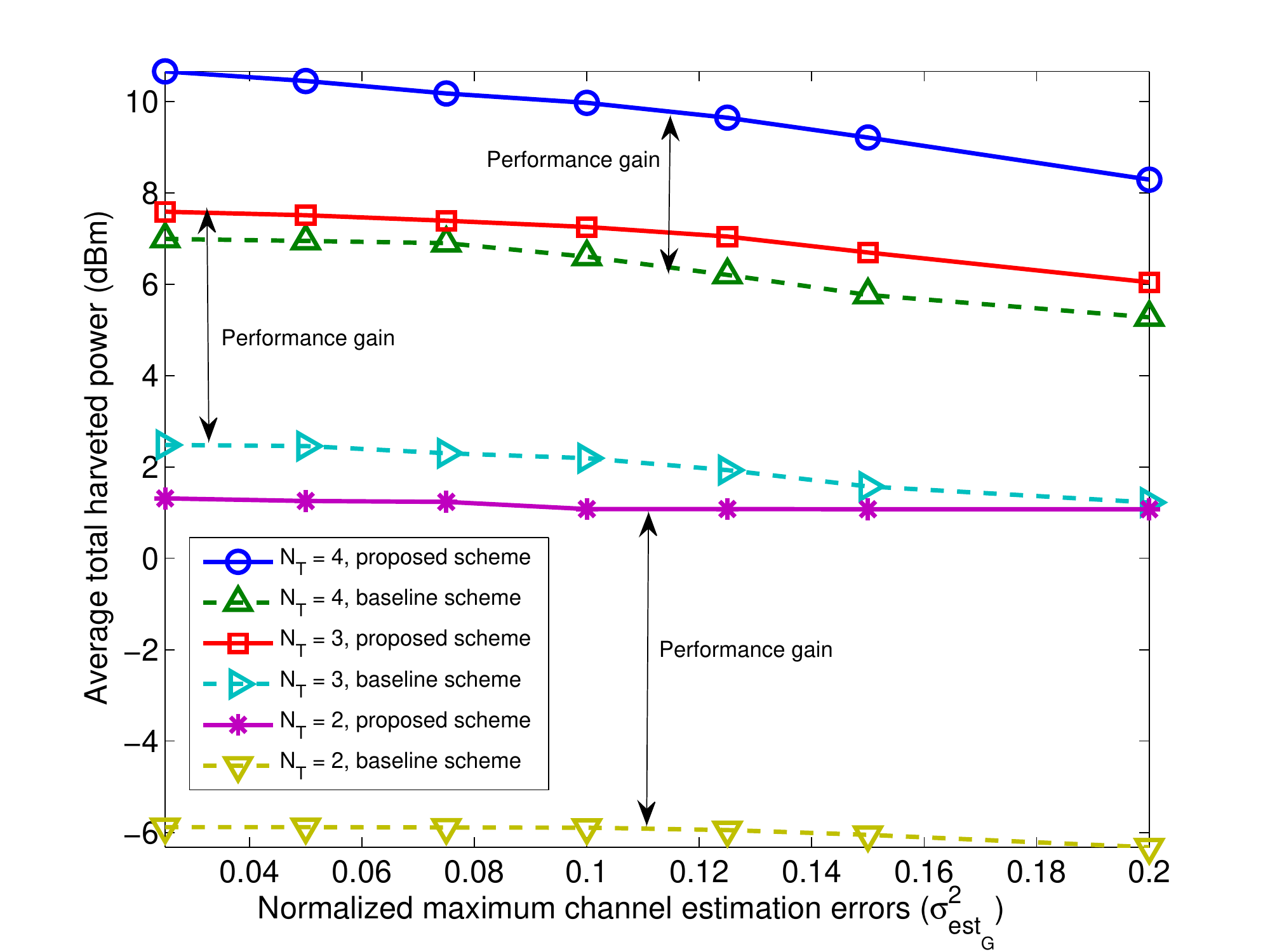}\vspace*{-2mm}
 \caption{Average total harvested power (dBm) versus  the normalized maximum channel estimation error. }
 \label{fig:hp_SINR}
 \centering
\includegraphics[width=3.2in]{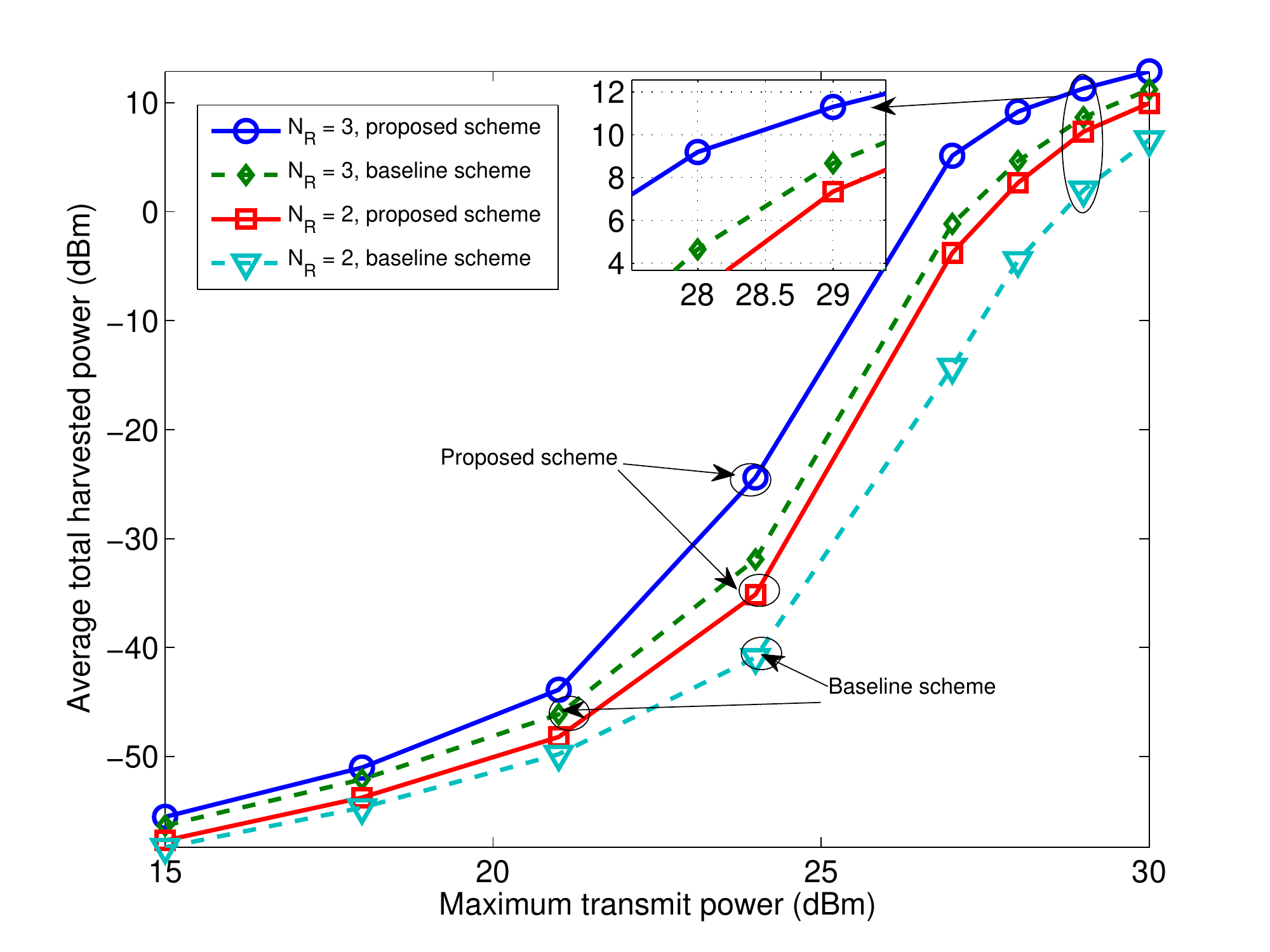}\vspace*{-2mm}
 \caption{Average total harvested power (dBm) versus  the maximum transmit power (dBm). }
 \label{fig:cap_SINR}\vspace*{-6mm}
\end{figure}

In Figure \ref{fig:hp_SINR}, we study the average total harvested power versus the maximum channel estimation error $\sigma_{\mathrm{est}_{G}}^2$, for different numbers of transmit antennas  and resource allocation schemes. The maximum transmit power is $P_{\max}=30$ dBm and $N_\mathrm{R}=2$. As can be observed, the total harvested power decreases with increasing $\sigma_{\mathrm{est}_{G}}^2$, since the CSI quality degrades with increasing $\sigma_{\mathrm{est}_{G}}^2$. In particular, for a larger value of $\sigma_{\mathrm{est}_{G}}^2$,  it is more difficult for the transmitter to steer the transmission towards the energy harvesting receivers accurately to improve the  efficiency of  wireless power transfer. On the other hand, the total harvested power in the system improves with increasing number of transmit antennas. This is because the  extra degrees of freedom introduced by additional transmit antennas can be exploited for a more efficient resource allocation. Furthermore, the proposed optimal scheme is able to fulfill the minimum required receive SINR in all considered scenarios,  despite the  imperfect CSI knowledge.

For comparison, we also show  the performance of a baseline scheme. For the baseline scheme, the resource allocation algorithm is designed for the conventional linear energy harvesting model  \cite{CN:Shannon_meets_tesla}--\cite{JR:Kwan_secure_imperfect}. Besides, the transmitter of the baseline scheme treats the estimated channel matrices $\mathbf{\hat G}_j,\forall j,$  as perfect CSI for resource allocation. Then, we optimize the power of $\mathbf{w},\mathbf{V}$ subject to the constraints in \eqref{eqn:TP_maximization}. It can be observed that the proposed optimal algorithm provides a substantial performance gain compared to the baseline scheme, particularly when the estimation errors are comparatively large. In fact,  the baseline scheme   may cause mismatches in resource allocation   since it does not account for the non-linear nature of the energy harvesting circuits.

Figure \ref{fig:cap_SINR} illustrates the average total harvested power versus the maximum transmit power for different number of receive antennas $N_{\mathrm{R}}$. The normalized maximum channel estimation error is $\sigma_{\mathrm{est}_{G}}^2=0.1$ and $N_{\mathrm{T}}=4$. As can be observed, the average total harvested power increases with the maximum transmit power non-linearly. In particular, when the maximum transmit power is small, e.g. $P_{\max}\le 21$ dBm, the total harvested power increases slowly with the transmit power.  In fact, most of the time, the received power at the energy harvesting receivers is insufficient for switching on the energy harvesting circuits. For a moderate transmit power level, e.g. $21\le P_{\max}\le 27$ dBm, the total harvested power increases rapidly respect to the transmit power.  However, when the transmit power is sufficiently large, e.g. $P_{\max}\ge 27$ dBm,  the total harvested power increases with the maximum transmit power with diminishing return. This is due to the fact that an exceedingly large transmit power causes saturation in some energy harvesting receivers. On the other hand,   when the number of antenna equipped at the energy harvesting receivers increases,  a significant energy harvesting gain can be achieved by the proposed optimal scheme. In fact, the extra receiver antennas act as additional energy collectors  which enables a more efficient energy transfer.

\section{Conclusions}\label{sect:conclusion}
In this paper, we  studied the resource allocation algorithm design for SWIPT based on a non-linear energy harvesting model and imperfect CSI.
 The algorithm design was formulated as a non-convex  optimization problem for the maximization of the total power transferred to the energy harvesting receivers.  The non-convex optimization problem was solved optimally with an iterative algorithm.  Numerical
results showed the potential gains in harvested power enabled  by  the
proposed optimization.

\section*{Appendix-Proof of Theorem \ref{thm:rankone}}\label{app:rankone}
If $\Rank(\mathbf{W})>1$ is obtained from \eqref{eqn:TP_maximization-transfomred}, we can construct an optimal rank-one solution as follows. For a given optimal $\bm\tau^*$ from the solution of the SDP relaxed version of \eqref{eqn:TP_maximization-transfomred}, we solve the following optimization problem:
\begin{eqnarray} \label{eqn:dummy_problem}
\underset{{\mathbf{W},\mathbf{V}\in \mathbb{H}^{N_{\mathrm{T}}}, \delta, \bm{\nu}}}{\mino}\,\, \hspace*{-1mm}&&\hspace*{-2mm} \Tr(\mathbf{W})\\
\hspace*{-1mm}\mathrm{subject\,\,to}\,\, &&\hspace*{-2mm}\mathrm{C1-C3, C6, C7,}\notag\\
\hspace*{-1mm}&&\hspace*{-2mm}\mathrm{C4}:\,\,\mathbf{S}_{\mathrm{C}_{4_j}}\Big(\mathbf{W},\mathbf{V}, \bm{\nu},\bm{\tau}^*\Big) \succeq \zero,\notag \forall j.
\end{eqnarray}
We note that the  optimal resource allocation policy obtained from \eqref{eqn:dummy_problem} is also an optimal resource allocation policy for the SDP relaxed version of \eqref{eqn:TP_maximization-transfomred}, since both problems have an identical feasible solution set.
Now, we aim to show that \eqref{eqn:dummy_problem} admits a rank-one beamforming matrix. To this end, we first need the Lagrangian of problem \eqref{eqn:dummy_problem} which is given by:
\begin{eqnarray}
L&=&\Tr(\mathbf{W})+\lambda(\Tr(\mathbf{W}+\mathbf{V})- P_{\mathrm{max}}) - \Tr(\mathbf{W}\mathbf{Y})\notag\\
&-&\sum_{j=1}^J\Tr(\mathbf{S}_{\mathrm{C}_{4_j}}\Big(\mathbf{W},\mathbf{V}, \bm{\nu},\bm{\tau}^*\Big)\mathbf{D}_{\mathrm{C}_{4_j}})\notag\\
&-&\Tr(\mathbf{S}_{\mathrm{C}_{2}}\Big(\mathbf{W},\mathbf{V},\delta\Big)\mathbf{D}_{\mathrm{C}_{2}})- \Tr(\mathbf{V}\mathbf{Z})+{\bm\Delta},
\end{eqnarray}
where $\lambda\geq 0$, $\mathbf{D}_{\mathrm{C}_{2}}\succeq\zero$, $\mathbf{Z}\succeq\zero$, $\mathbf{D}_{\mathrm{C}_{4_j}}\succeq\zero,\forall j\in\{1,\ldots,J\}$, $\mathbf{Y}\succeq\zero$, are the dual variables for constraints C1--C4, and C6, respectively.  $\bm\Delta$ is a collection of variables and constants that are not relevant to the proof.

Now, we focus on those Karush-Kuhn-Tucker (KKT) conditions which are needed for the proof:
\begin{eqnarray}
&&\hspace*{-6mm}\mathbf{Y}^*,\mathbf{V}^*,\mathbf{D}_{\mathrm{C}_{2}}^*,\mathbf{D}_{\mathrm{C}_{4_j}}^*\succeq \zero,\quad\lambda^*\ge0,\\
&&\hspace*{-6mm} \mathbf{Y^*W^*}=\zero,\quad \label{eqn:KKT-complementarity}  \mathbf{Q^*V^*}=\zero, \label{eqn:KKT-complementarity}\\
\label{eqn:KKT_Y1}
&&\hspace*{-6mm}\mathbf{Y^*}\hspace*{-0.5mm}=\hspace*{-0.5mm}(1\hspace*{-0.5mm}+\hspace*{-0.5mm}\lambda^*)\mathbf{I}_{N_\mathrm{T}}\hspace*{-0.5mm}-\hspace*{-0.5mm}\mathbf{U}_{\mathbf{\hat h}}\mathbf{D}_{\mathrm{C}_{2}}\mathbf{U}_{\mathbf{\hat h}}^H\hspace*{-0.5mm}-\hspace*{-0.5mm}{\bm\Xi} \\
\label{eqn:KKT_Y2}&&\hspace*{-6mm}\mathbf{Q^*}= \lambda^*\mathbf{I}_{N_\mathrm{T}}+\Gamma_{\mathrm{req}}\mathbf{U}_{\mathbf{\hat h}}\mathbf{D}_{\mathrm{C}_{2}}\mathbf{U}_{\mathbf{\hat h}}^H-{\bm\Xi},\\ \label{eqn:C2_complementarity}
&&\hspace*{-6mm}\mathbf{S}_{\mathrm{C}_{2}}\Big(\mathbf{W},\mathbf{V},\delta\Big)\mathbf{D}_{\mathrm{C}_{2}}=\zero,
\end{eqnarray}
where $ {\bm\Xi}=\sum_{j=1}^J\sum_{l=1}^{N_{\mathrm{R}}} \Big[\hspace*{-0.5mm}
\mathbf{U}_{\widetilde{\mathbf{g}}_j}\mathbf{D}_{\mathrm{C}_{4_j}}\hspace*{-0.5mm}\mathbf{U}_{\widetilde{\mathbf{g}}_j}^H \hspace*{-0.5mm}\Big]_{a:b,c:d}, a=(l-1)N_{\mathrm{T}}+1,b=l N_{\mathrm{T}},c=(l-1)N_{\mathrm{T}}+1,$ and $d=l N_{\mathrm{T}}$. The optimal
primal and dual variables of the SDP relaxed version are denoted by the corresponding
variables with an asterisk superscript.

Then, we follow a similar approach as \cite{JR:SWIPT_imperfect_CSI} to show that $\Rank(\mathbf{W}^*)=1$. Subtracting \eqref{eqn:KKT_Y2} from \eqref{eqn:KKT_Y1} yields:
\begin{eqnarray}\label{eqn:temp_eq}
\mathbf{Y}^*+ (1+\Gamma_{\mathrm{req}})\mathbf{U}_{\mathbf{\hat h}}\mathbf{D}_{\mathrm{C}_{2}}\mathbf{U}_{\mathbf{\hat h}}^H= \mathbf{Q}^*+ \mathbf{I}_{N_\mathrm{T}}.
\end{eqnarray}
Next, we multiply the both sides of \eqref{eqn:temp_eq} by $\mathbf{W}^*$ leading to
\begin{eqnarray}\label{eqn:temp_eq2}
\mathbf{W}^* (1+\Gamma_{\mathrm{req}})\mathbf{U}_{\mathbf{\hat h}}\mathbf{D}_{\mathrm{C}_{2}}\mathbf{U}_{\mathbf{\hat h}}^H= \mathbf{W}^*(\mathbf{Q}^*+ \mathbf{I}_{N_\mathrm{T}}).
\end{eqnarray}
From \eqref{eqn:temp_eq2}, we can deduce that
\begin{eqnarray}\label{rank_inequality}
\hspace*{-1mm}\Rank(\mathbf{W}^*)\hspace*{-3.5mm}&=&\hspace*{-2mm}\Rank(\mathbf{W}^*(1+\Gamma_{\mathrm{req}})\mathbf{U}_{\mathbf{\hat h}}\mathbf{D}_{\mathrm{C}_{2}}\mathbf{U}_{\mathbf{\hat h}}^H)\\
\hspace*{-1mm}\hspace*{-3.5mm}&\leq&\hspace*{-2mm} \min\{\hspace*{-0.5mm}\Rank(\mathbf{W}^*),\hspace*{-0.5mm} \Rank((1\hspace*{-0.5mm}+\hspace*{-0.5mm}\Gamma_{\mathrm{req}})\mathbf{U}_{\mathbf{\hat h}}\mathbf{D}_{\mathrm{C}_{2}}\mathbf{U}_{\mathbf{\hat h}}^H)\hspace*{-0.5mm}\}.\notag
\end{eqnarray}
Therefore, if $\Rank(\mathbf{U}_{\mathbf{\hat h}}\mathbf{D}_{\mathrm{C}_{2}}\mathbf{U}_{\mathbf{\hat h}}^H)\leq 1$, then $\Rank(\mathbf{W}^*)\leq 1$. To show $\Rank(\mathbf{U}_{\mathbf{\hat h}}\mathbf{D}_{\mathrm{C}_{2}}\mathbf{U}_{\mathbf{\hat h}}^H)\leq 1$, we pre-multiply and post-multiply \eqref{eqn:C2_complementarity} by $[\mathbf{I}_{N_{\mathrm{T}}}\,\zero]$ and $\mathbf{U}_{\widetilde{\mathbf{g}}_j}^H$, respectively. After some mathematical manipulations, we have the following equality:
\begin{eqnarray}\label{rank_inequality2}
&&(\delta\mathbf{I}_{N_{\mathrm{T}}}\hspace*{-0.5mm}+\hspace*{-0.5mm}(\mathbf{W}^*\hspace*{-0.5mm}-\hspace*{-0.5mm}(1\hspace*{-0.5mm}+\hspace*{-0.5mm}\Gamma_{\mathrm{req}})\mathbf{V}^*))\mathbf{U}_{\mathbf{\hat h}}\mathbf{D}_{\mathrm{C}_{2}}\mathbf{U}_{\mathbf{\hat h}}^H \notag\\
&=&\hspace*{-0.5mm}\delta[\zero\,\mathbf{\hat h}]\mathbf{D}_{\mathrm{C}_{2}}\mathbf{U}_{\mathbf{\hat h}}^H.
\end{eqnarray}
Besides, it can be shown that $(\delta\mathbf{I}_{N_{\mathrm{T}}}\hspace*{-0.5mm}+\hspace*{-0.5mm}(\mathbf{W}^*\hspace*{-0.5mm}-\hspace*{-0.5mm}(1\hspace*{-0.5mm}+\hspace*{-0.5mm}\Gamma_{\mathrm{req}})\mathbf{V}^*))\succ\zero$ and $\delta>0$ hold at the optimal solution such that the dual optimal solution is bounded from above. Therefore,  we have
\begin{eqnarray}\label{rank_inequality3}
&&\Rank(\mathbf{U}_{\mathbf{\hat h}}\mathbf{D}_{\mathrm{C}_{2}}\mathbf{U}_{\mathbf{\hat h}}^H)\\
&=&\notag\Rank(\delta[\zero\,\mathbf{\hat h}]\mathbf{D}_{\mathrm{C}_{2}}\mathbf{U}_{\mathbf{\hat h}}^H)\leq\Rank([\zero\, \hat {\mathbf{h}}])   \leq 1.
\end{eqnarray}
By combining \eqref{rank_inequality} and \eqref{rank_inequality3}, we can conclude that $\Rank(\mathbf{W}^* )\leq 1$. On the other hand, since $\Gamma_{\mathrm{req}}>0$, $\mathbf{W}^*\neq \zero$ holds and $\Rank(\mathbf{W}^*)=1$. \qed

\end{document}